\documentclass[showpacs,twocolumn,prl]{revtex4}
\usepackage{times}
\usepackage{amssymb}
\usepackage{amsmath}
\usepackage{mathrsfs}
\usepackage{graphicx}
\usepackage{epsfig}
\usepackage{dcolumn}
\usepackage{color}
\usepackage{bm}
\begin{document}
\title{Vacuum-Induced Coherence in Ultracold Photoassociative Ro-Vibrational Excitations}
\author{Sumanta Das$^{1,\footnote{}}
 $, Arpita Rakshit$^2$ and Bimalendu Deb$^{2,3}$}
\affiliation{$^1$1RC 58/6 Uttar Raghunathpur, Teghoria Kolkata 700059, India \\ 
$^2$Department of Materials Science, and 
$^4$Raman Center for Atomic, Molecular and Optical Sciences, Indian Association
for the Cultivation of Science,
Jadavpur, Kolkata 700032, India.}

\begin{abstract}
We show that coherence between two excited ro-vibrational states belonging to the same molecular electronic configuration arises quite naturally due to their interaction with electromagnetic vacuum. For initial preparation of a molecule in the desired ro-vibrational states, we propose to employ the method of ultracold photoassociation. Spontaneous decay of the excited molecule then gives rise to vacuum induced coherence between the excited ro-vibrational states.  We demonstrate theoretically an interesting interplay of effects due to vacuum induced coherence and photoassociation. We apply our theory to photoassociation of bosonic Ytterbium ($^{174}$Yb) atoms which appear to be a promising system for exploring such interplay. The effects discussed here can be important for controlling decoherence and dissipation in molecular systems.
\end{abstract}
\maketitle
\footnotetext[0]{Current address: Max-Planck-Institut f\''ur Kernphysic, Heidelberg,
Germany}
Quantum coherences and interference  form the basis of many fascinating phenomena in atomic 
and molecular systems \cite{mandel,ficek}. Its applications in spectroscopy, metrology, photovoltaics 
and information sciences have been most remarkable. One of the famous examples of  quantum interference is the Fano effect \cite{fano,miroshnichenko} that arises from interference of two competing optical transition pathways involving discrete and continuum states. 
Another notable coherence phenomenon is the vacuum-induced coherence (VIC) 
which arises due to quantum interference between two pathways of spontaneous emissions \cite{agarwal}. It is known that under appropriate conditions, VIC can lead to population trapping in excited states \cite{agarwal, ficek}. This can be utilized in manipulating environment induced relaxation processes in a wide variety of systems such as atoms, ions, molecules, quantum dots \cite{agarwal, gauthier, scully1, garraway, zhou, dutta, das08, xia},  and has also been found to be effective against decoherence in quantum information processing \cite{Das10}. One of the key conditions for VIC to occur is the nonorthogonality of the dipole moments of two spontaneous transitions. For atomic systems nonorthogonality is a stringent condition to achieve. Possible realization of VIC for an excited atom interacting with an anisotropic vacuum \cite{agarwal1, yang, evan} and utilizing the $j = 1/2 \rightarrow j = 1/2$ transition in $^{198}$Hg$^+$ and $^{139}$Ba$^+$ ions have been suggested \cite{kiffner, das}.  Recently, a proof-of-principle experiment verifying its presence has been performed in quantum dots \cite{dutt}. In-spite of these attempts, a clear signature of VIC in atomic systems is yet to be obtained.
\begin{figure}
\includegraphics[width=\columnwidth]{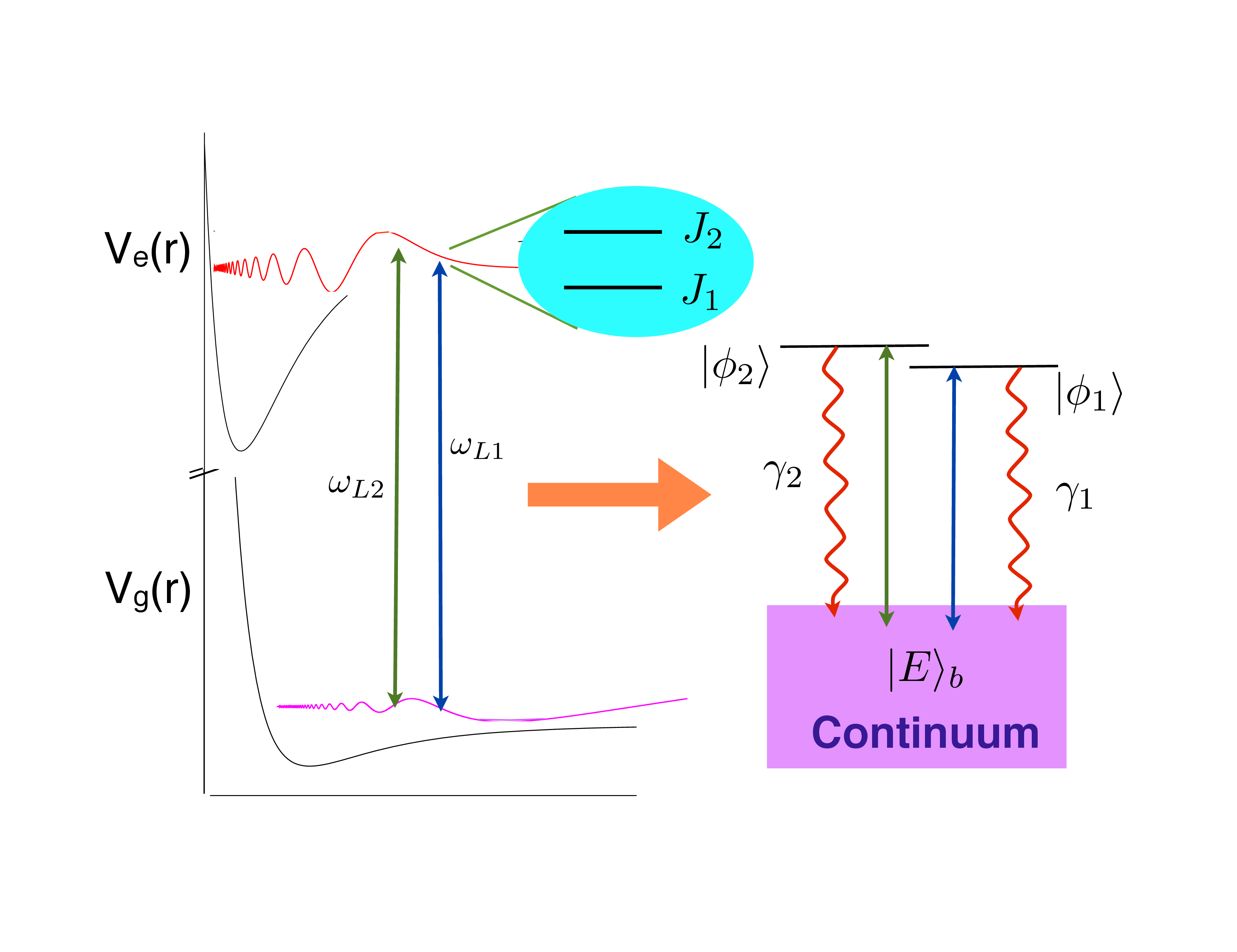}
\caption{(Color online) A schematic diagram for creation of coherence in two excited 
ro-vibrational levels by  photoassociation. The two-atom continuum $|E\rangle_b$ 
is coupled to two ro-vibrational levels $|{\it{v}},J_1\rangle (|\phi_1\rangle)$ 
and $|{\it{v}},J_2\rangle (|\phi_2\rangle)$ of an excited molecular electronic 
state via two lasers of frequencies $\omega_{L1}$ and $\omega_{L2}$, respectively. 
$\gamma_1$ and $\gamma_2$ are the spontaneous decay rates at which  $|\phi_1\rangle$ 
and $|\phi_2\rangle$ decay to the continuum. Coherence arises between the two excited 
molecular states from the quantum interference of two spontaneous 
decay pathways $ |\phi_1\rangle \rightarrow |E\rangle_b $ and  $ |\phi_2\rangle \rightarrow |E\rangle_b $. }
\end{figure}

In this Rapid Communication we show that unlike atoms or ions, VIC arises quite naturally in molecular systems.  
This can be attributed to the quantum interference of  spontaneous emission pathways 
from two ro-vibrational levels of an excited molecular electronic state to a ground molecular state. 
In this case, the required nonorthogonality condition is satisfied naturally as the two excited 
states belong to the same molecular electronic configuration and differ only in rotational or 
vibrational quantum numbers. With the tremendous progress in high precision photoassociation 
(PA) spectroscopy \cite{weiner,jones}, low lying rotational levels can be selectively populated 
in a molecular excited state. This occurs due to PA transitions from the collisional continuum 
of two ground state cold atoms. VIC will be significant in such an atom-molecule interface system 
provided (i) there is no hyperfine interaction in the atoms that are photoassociated, 
(ii) there is no bound state close to the dissociation continuum of the
ground molecular state and (iii) excited molecular levels have long life time. 

However, to our knowledge, the possibility of VIC in such PA systems has not been 
addressed so far. As such, we propose here a novel PA scheme for realization of VIC 
in an atom-molecule system.  We present results on the interplay of VIC and the effects 
induced by PA lasers such as dressing of the continuum. We show that the life time of the excited states can be controlled by appropriate manipulation 
of this interplay. Moreover, our results predict that for an optimum detuning and intensity of PA lasers, 
the interplay between VIC and PA can lead to coherent population trapping in the excited states.

The basic idea of our scheme is depicted in Fig. 1. We consider as our model 
a system of two excited molecular ro-vibrational levels 
$|{\it{v}},J_1\rangle (|\phi_1\rangle)$ and $|{\it{v}},J_2\rangle (|\phi_2\rangle)$ 
coupled to the two-atom ground continuum $|E\rangle_b$ via lasers. Initially either $|\phi_1\rangle$ or  $|\phi_2\rangle$ is populated or  partially both are populated via photoassociation of cold atoms using two lasers $L_1$ and $L_2$ of frequencies $\omega_{L1}$ and $\omega_{L2}$, tuned near  $\mid E \rangle_b \rightarrow \mid \phi_1\rangle$ and 
$\mid E \rangle_b \rightarrow \mid \phi_2\rangle$ transitions, respectively. 
Both the excited levels $|\phi_{1}\rangle$ and $|\phi_{2}\rangle$ decay 
spontaneously to the same ground continuum with decay rates $\gamma_{1}$ and $\gamma_{2}$, respectively.  
The Hamiltonian governing the dynamics of this system can be written as 
$\mathcal{H} = \mathcal{H}_{c} + \mathcal{H}_{inc}$, 
where $\mathcal{H}_{c}$ is the coherent part involving PA couplings and is given by, 
\begin{eqnarray}
\mathcal{H}_{c} &=& \sum_{n=1}^2\hbar\omega_{bn}\mid \phi_n\rangle\langle\phi_n\mid + 
\int{E' \mid E' \rangle_{b}\hspace{1mm}_{b}\langle E' \mid dE' }\nonumber\\ 
&+&\int{\sum_{n=1}^2  \left\lbrace \Lambda_{nE'} e^{-i\omega_{Ln}t} \hat{S}_{E'}^\dagger  + 
{\rm H.C.}\right\rbrace dE'}.
\end{eqnarray}
Here $\hbar\omega_{bn}$ are the binding energies of the bound states $|\phi_{n}\rangle ($n$ = 1,2)$; $|E'\rangle_b$ is 
the bare continuum state and $\Lambda_{nE'} = \langle \phi_n \mid \vec{\mathscr{D}}_n\cdot\vec{E}_{Ln} 
\mid E'\rangle_b$ is the laser coupling for the transition from the $n$-th 
bound state to the bare continuum $|E'\rangle_b$. The vectors $\vec{\mathscr{D}}_{n}$ and $\vec{E}_{Ln}$ are the dipole 
moment and electric field of the laser associated with the $n$-th transition, respectively. 
The operator $\hat{S}_{E'}^\dagger = |\phi_n\rangle _b\langle E' \mid$ is a raising
operator. $\mathcal{H}_{c}$ is exactly diagonalizable \cite{bdeb1, bdeb2} in the spirit of Fano's theory \cite{fano}. 
The eigen state of $\mathcal{H}_{c}$ is the dressed continuum  expressed as
\begin{eqnarray}
\mid E \rangle_{d} = A_{2E}|\phi_2\rangle + A_{1E}|\phi_1\rangle + \int{C_{E^{\prime}}(E) 
\mid E^{\prime}\rangle_b dE^{\prime}}
\end{eqnarray}
with the normalization condition $\langle E^{\prime\prime}|E\rangle = \delta(E-E^{\prime\prime})$. 
Here
$A_{nE} =\Lambda_{nE}(q\Gamma_{n^\prime}/2 +  \delta_{n^{\prime}E})/\hbar[(\delta_{1E} +
i\Gamma_1/2)(\delta_{2E} +i\Gamma_2/2)-\Gamma_1\Gamma_2(q-i)^2/4]$
with $n' \neq n$ and
$C_{E^{\prime}}(E) =\delta(E-E^{\prime})+ (A_{1E}\Lambda_{1E^{\prime}}+A_{2E}\Lambda_{2E^{\prime}})/
(E-E^{\prime})$.
The term $\Gamma_n(E) = 2\pi|\Lambda_{nE}|^2/\hbar$, is the stimulated line width 
of the $n$-th bound state due to continuum-bound laser coupling and $\delta_{nE} = E/\hbar - \delta_n$ with detuning $\delta_n =\omega_{bn}-\omega_{Ln}$. Here $q = V_{12}/(\pi\Lambda_{1E}\Lambda_{2E})$ is analogous to the well-known Fano's $q$ parameter \cite{fano} with $ V_{12} =  \mathscr{P} \int d E' \Lambda_{1E'} \Lambda_{2E'}/(E - E')$ where `$\mathscr{P}$' stands for principal value.

The incoherent part of the Hamiltonian $\mathcal{H}_{inc}$ describes the interaction of 
vacuum field with the system and is given by,
\begin{eqnarray}
H_{inc} = \int{dE^{\prime}\left[\sum_{n=1}^2\sum_{\kappa,\sigma}g_{n,\sigma}(E^{\prime},\kappa)
\hat{S}_{E^{\prime}}^\dagger \hat{a}_{\kappa,\sigma}e^{-i\omega_\kappa t}  + {\rm H.c.}\right]}
\end{eqnarray}
where $\hat{a}_{\kappa,\sigma}$ is the annihilation operator of the vacuum field 
$\vec{E}_{vac}$ and $g_{n,\sigma}(E^{\prime},\kappa) = -\langle \phi_n \mid \vec{\mathscr{D}}_i\cdot
\vec{E}_{vac}(\kappa)\mid E^{\prime} \rangle_{b}$ is the dipole coupling with 
$\vec{E}_{vac}(\kappa) = \left(\sqrt{\hbar\omega_{\kappa}/2\epsilon_{0}V}\right)\vec{\varepsilon}_{\sigma}$,
$\kappa$ being the wave number, $\sigma$ the polarization of the field and $\sqrt{\hbar\omega_{\kappa}/2\epsilon_{0}V}$ the 
amplitude of the vacuum field.
Let the joint state of the system-reservoir at a time $t$ be expressed as,
\begin{eqnarray}
\mid \Psi (t)\rangle & = &\sum_n a_n(t)\mid \phi_n,\{0\}\rangle \nonumber\\
& + & \int{dE^{\prime}\sum_{\kappa,\sigma}b_{E^{\prime},\kappa,\sigma}}(t)\mid E^{\prime}\rangle_b \mid\{1_{\kappa,\sigma}\}\rangle
\end{eqnarray}
where $a_{n}$ and $b_{E^{\prime},\kappa,\sigma}$ are the  amplitudes of $n$-th excited state and ground continuum, respectively. The state $|\phi_{n}, \{0\}\rangle$  corresponds to molecular excited state with field in vacuum and $|E^{\prime}\rangle_{b}|\{1_{\kappa\sigma}\}\rangle$ refers to (ground) bare continuum state with energy $E'$ and one photon 
in mode $\kappa$ of polarization $\sigma$. Using the standard Wigner-Weisskopf approach \cite{ficek}, after a long algebra we obtain,
\begin{eqnarray}
\dot{\tilde{a}}_1 &=& \left(- {\cal{G}}_{1} - i\delta_{12} \right) \tilde{a}_{1} - {\cal{G}}_{12} \tilde{a}_{2}\label{e}\\
\dot{\tilde{a}}_2 &=& - {\cal{G}}_{2} \tilde{a}_{2} - {\cal{G}}_{21}\tilde{a}_{1}\label{f}
\end{eqnarray}
where $\tilde{a}_{n}$ is the modified amplitude related to $a_n$ by transformations,
$\tilde{a}_1=a_1 \exp{[i(\tilde{\omega}_1 - \delta_{12})t]}$ and $\tilde{a}_2=a_2 \exp{[i\tilde{\omega}_2t]}$. The  dressed frequency  $\tilde{\omega}_{n}$ and the detuning ${\delta}_{12}$ in the above expression are given by 
$\hbar\tilde{\omega}_n = \int{|A_{nE}|^2 E dE}$ and  $\delta_{12} = (\omega_{L1} -\omega_{L2})
 + (\tilde{\omega}_1-\tilde{\omega}_2)$. The ${\cal{G}}_{n}$ is the decay constant of the $n$-th bound state, given by
\begin{eqnarray} 
{\cal{G}}_{n} &=& \frac{1}{\hbar^2} \int{dE' \int {dt'\sum_{\kappa,\sigma} |g_{n \sigma}(E',\kappa)|^2}}\nonumber\\
& &{{\exp{\left[i\left(\omega_{Ln}+\tilde{\omega}_n - \frac{E'}{\hbar} -\omega_{\kappa}\right)(t-t')\right]}}}
\end{eqnarray} 
A key feature of Eqs. $(\ref{e}-\ref{f})$ is the coupling between the amplitudes via the cross damping term  
\begin{eqnarray}
{\cal{G}}_{nn'} \simeq \frac{1}{\hbar^2}\int{dE' \int{dt'\sum_{\kappa,\sigma} g_{n \sigma}(E',\kappa)g^*_{n' \sigma}(E',\kappa)}}\nonumber\\{{\exp\left[{i\left(\omega_{Ln}+\tilde{\omega}_n -\frac{E'}{\hbar}-\omega_{\kappa}\right)(t-t')}\right]}}
\end{eqnarray} 
where $n \neq n'$. For simplicity in writing the above equation we have assumed $(\omega_{L1}+\tilde{\omega}_1) \simeq (\omega_{L2}+\tilde{\omega}_2)$. It is important to understand that ${\cal{G}}_{12}$ arises due to quantum interference of the spontaneous emission pathways resulting in VIC between  the excited states amplitudes.  Summing over the vacuum modes and then carrying out the time integral under Born Markov approximation, we finally obtain  ${\cal{G}}_n = (3\pi\epsilon_0\hbar c^3)^{-1}\int {dE^{\prime} (\omega_{Ln} + \tilde{\omega}_n -E'/\hbar)^3}{ |\langle\phi_n|\vec{\mathscr{D}}_{n}|E^{\prime}\rangle_b|^2}$  and ${\cal{G}}_{12}  = {\cal{G}}_{21}  = (3\epsilon_0\hbar c^3 \pi)^{-1}\int {dE^{\prime} \langle\phi_1|\vec{\mathscr{D}}_{1}|E^{\prime}\rangle \langle E^{\prime}|\vec{\mathscr{D}}_{2}|\phi_2 \rangle}$ ${(\omega_{L1}+\tilde{\omega}_1-E'/\hbar)^{3/2}(\omega_{L2}+\tilde{\omega}_{2} -E'/\hbar)^{3/2}}$.
\begin{figure}
\includegraphics[width=\columnwidth]{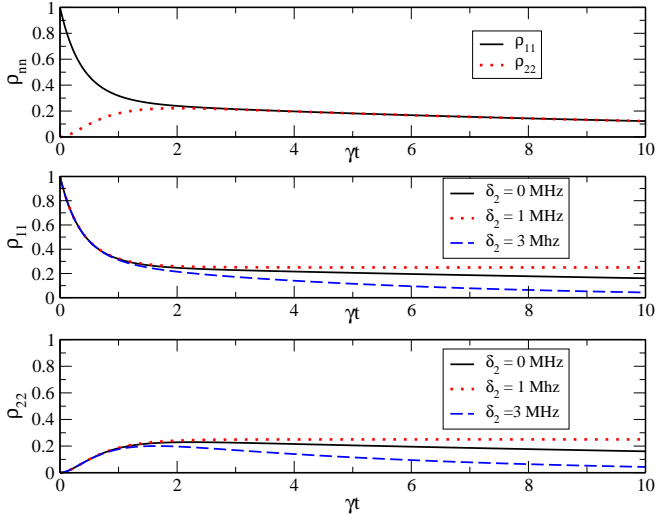}
\caption{(Color online) Uppermost Panel shows the plots of $\rho_{11}$  and $\rho_{22}$ in the absence of laser which is the case of normal VIC. Both $\rho_{11}$  and $\rho_{22}$ go to zero in the long time limit. In the middle and lower panel, the excited state population $\rho_{11}$  and $\rho_{22}$ are plotted as a function of dimensionless time $\gamma t$ for different values of the detuning of the second laser, $\delta_2$, keeping $\delta_1 = 0$. At $\delta_2 = 1$ MHz, the population is trapped between the two excited states. The intensities of $L_1$ and $L_2$ are 50 mW cm$^{-2}$ and 0.1 mW cm$^{-2}$, respectively.}
\label{fig:2}
\end{figure}
Solving the coupled Eqs. (5) and (6) analytically, we obtain
\begin{eqnarray}
\label{sol1}
\tilde{a}_1 (t)& = & c_{1-} e^{z_- t} + c_{1+} e^{z_+ t}\\
\label{sol2}
\tilde{a}_2 (t)& = & c_{2-} e^{z_- t} + c_{2+} e^{z_+ t}
\end{eqnarray}
where, $ z_{\pm} = \frac{1}{2}[-i\delta_{12} - {\cal{G}}_+ \pm \Omega]$ and 
$c_{1\pm} = [\pm(-i\delta_{12}-{\cal{G}}_-\pm\Omega)\tilde{a}_{1}(0)\mp2{\cal{G}}_{12}\tilde{a}_{2}(0)]/(2\Omega)$ 
and $c_{2\pm} = [\pm(i\delta_{12} + {\cal{G}}_-\pm\Omega)
\tilde{a}_{2}(0)\mp2{\cal{G}}_{12}\tilde{a}_{2}(0)]/(2\Omega)$ 
with $\Omega = \sqrt{({\cal{G}}_- +i\delta_{12})^2 + 4{\cal{G}}_{12}^2}$, ${\cal{G}}_- ={\cal{G}}_1 - {\cal{G}}_2$ and ${\cal{G}}_+ ={\cal{G}}_1 + {\cal{G}}_2$.

In the limit when laser intensities going to zero (weak coupling), we find $\tilde{\omega_n} \rightarrow (\omega_{bn} 
-\omega_{Ln})$, thus ${\cal{G}}_n$ reduces to the usual damping constant $\gamma_{n}$. Moreover, for low energy we have
\begin{eqnarray}
{\cal{G}}_{12} \simeq \gamma_{12} =\frac{\sqrt{(\omega_{b1}\omega_{b2})^3 }}{(3\pi\epsilon_0\hbar c^3)} \langle \phi_1\mid \vec{\mathscr{D}_1} \cdotp \vec{\mathscr{D}_2}\mid \phi_2 \rangle.
\end{eqnarray}
Thus in the absence of lasers, the model reduces to normal VIC case in V-type system \cite{ficek}. The above 
equation shows that $\gamma_{12}$ vanishes if the molecular transition dipole moments $\vec{\mathscr{D}_1}$ and  
$\vec{\mathscr{D}_2}$  are orthogonal. In our model $\vec{\mathscr{D}_1}$ and $\vec{\mathscr{D}_2}$ are the 
transition dipole moments between the same ground and excited electronic states, therefore they are essentially
nonorthogonal.
\begin{figure}
\includegraphics[width=\columnwidth]{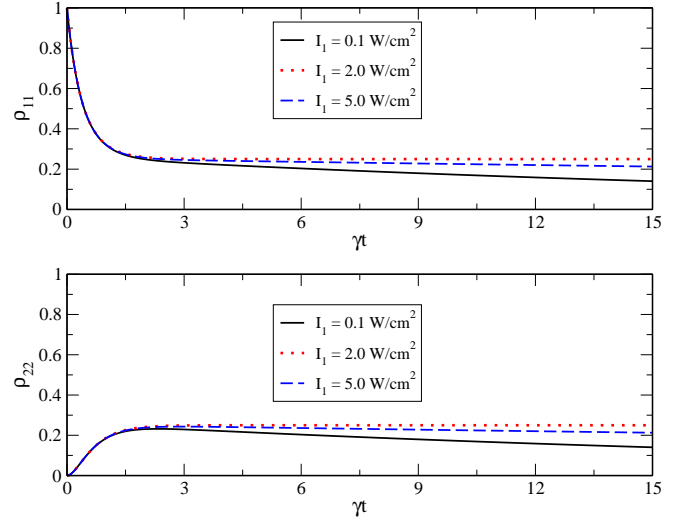}
\caption{(Color online) Same as in Fig. 2, but for different values of intensity of first laser, $I_1$, keeping the 
intensity of second laser fixed at 0.1 mW/cm$^{-2}$, keeping $\delta_1 = \delta_2 = 0$. The population 
is trapped at excited states for intensity of 2 W/cm$^{-2}$. }
\label{fig:3}
\end{figure}

The excited state populations  $\rho_{11} = |\tilde{a}_1|^2$, $\rho_{22} = |\tilde{a}_2|^2$  and the coherence $\rho_{12} = \tilde{a}_1 \tilde{a}_2^{'*}$ can be obtained from Eqs. (\ref{sol1}) and (\ref{sol2}). Explicitly 
\begin{eqnarray}
\rho_{nn} (t) &=& \frac{e^{-2{\cal{G}}_1 t}}{4\Omega^2}\left[ 2 \left\lbrace A \Omega^2 -B\delta_{12}^2\right\rbrace\rho_{nn} (0)\right.  \nonumber\\& & \left. - 8B{\cal{G}}_{12}^2\rho_{n'n'}(0)   +8B{\cal{G}}_{12}\delta_{12} \rm{Im}[\rho_{12}(0)]  \right.  \nonumber\\& & \left.  -8 {\cal{G}}_{12}\Omega \rm{Re}[\rho_{12}(0)] \sinh (\Omega t)\right] ,
\end{eqnarray}
where $n' \neq n$, $A = \left[1 + \cosh(\Omega t)\right]$ and $ B = \left[1 - \cosh (\Omega t)\right]$ and we have considered ${\cal{G}}_1 = {\cal{G}}_2$. At $4{\cal{G}}_1^2 = [4{\cal{G}}_{12}^2-\delta_{12}^2]$, it follows from above equation that $\rho_{11}$ and $\rho_{22}$ become time-independent in long time limit meaning  coherent population trapping in the excited states.  When $\delta_{12} = 0$,  $\rho_{11} (t \rightarrow \infty) = \rho_{22}(t \rightarrow \infty) = \left[ \rho_{11}(0) + \rho_{22}(0) -2Re[\rho_{12}(0)]\right]/4$ become exactly same as normal VIC case \cite{ficek}. It is worthwhile to emphasize that the results given in Eqs. (9), (10) and (12) are general because they are applicable to any PA coupling regime. 

For experimental realization of VIC, our model can be applied to the spin forbidden intercombination 
transition $^1S_0 - ^3P_1$ of  bosonic $^{174}$Yb \cite{takahashi1,takahashi2,takahashi3} which has no hyperfine interaction. 
The only molecular ground electronic state of $^{174}$Yb is $^1\Sigma_g$ which corresponds to  
$^1S_0 + ^1S_0$ at long separation and represents the only bare continuum $|E\rangle_{b}$ of our model and . 
The excited states $|\phi_{n}\rangle$ can be chosen as ro-vibrational levels in long range  $O_u^+$ state that can be populated by PA. For  illustration,  we specifically consider excited ro-vibrational levels $|\phi_{1}\rangle = |v = 118, J = 1\rangle$ and $ |\phi_{2}\rangle = |v = 118, J =3\rangle $ \cite{takahashi2}. According to the selection rules of continuum-bound transitions, the minimum partial wave $(l)$ that be coupled to  $ |\phi_{2}\rangle$ by PA is $d$ wave $(l = 2)$. Usually at ultracold temperatures, $d$ wave scattering amplitude becomes insignificant due to large centrifugal barrier. But ground state scattering properties of $^{174}$Yb are exceptional in the sense that it exhibits a prominent $d$-wave shape resonance at temperatures as low as 25 $\mu$K \cite{takahashi1,takahashi5}. 
  
\begin{figure}
\includegraphics[width=\columnwidth]{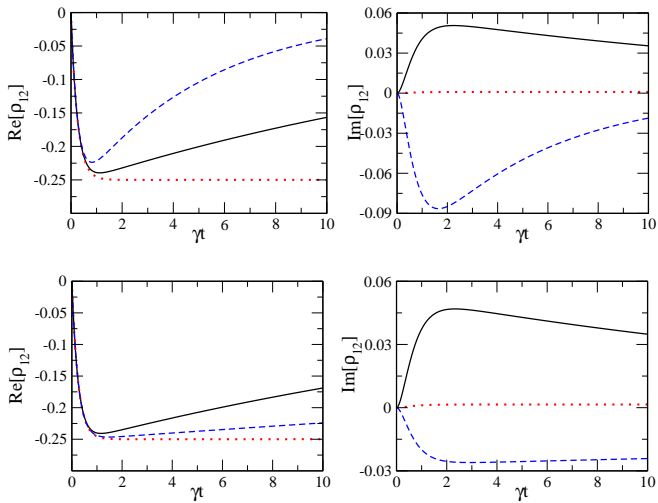}
\caption{(Color online) Plot of real and imaginary part of $\rho_{12}$   against $ \gamma t $ for different detunings of second laser (two upper subplots ) and different intensities of the first laser (two lower subplots). All other parameters of the upper subplots is same as in Fig. 2 while the other parameters of lower subplots are the same as in Fig. 3. In the two upper subplots, the detuning  $\delta_2 = 0$ (black solid line), $\delta_2 = 1$ MHz (red dotted line) and $\delta_2 = 3$ MHz (blue dashed lines) while in the lower  two subplots the intensity $I_1 = 0.1$ W cm$^{-2}$(black solid line), $I_1 = 2.0$ W cm$^{-2}$ (red dotted line) and $I_1 = 5.0$ W cm$^{-2}$ (blue dashed lines)}
\label{fig:4}
\end{figure}

We now discuss our numerical results.   In Fig. 2, we show the dynamical behavior of populations $\rho_{11}$ and $\rho_{22}$ as a function of scaled time $\gamma t$. The upper most panel shows populations in the absence of the lasers assuming  $\gamma_1 = \gamma_2 = \gamma_{12} =\gamma = 2.29$ MHz \cite{takahashi4}. The short time dynamics clearly shows exchange of population between $|\phi_1\rangle$ and $|\phi_2 \rangle$ due to VIC. In the lower two panels, we plot $\rho_{11}$ and $\rho_{22}$ of Eq. (12) for different values of $\delta_2$, keeping $\delta_1$ fixed. We find that, as $\delta_2$  increases upto an optimum frequency, the lifetime of both the excited levels also increases. Then at an optimum frequency , the population gets trapped in the excited state. For the parameters of Fig. 2, this optimum frequency is found to be 1 MHZ.  When $\delta_2$ increases beyond the optimum value, population falls off.  Since the value of dressed frequency $\tilde{\omega}_n$ depends upon the PA laser intensities, 
we expect the dynamics to be intensity dependent. Hence by varying 
the laser intensity of one of the PA lasers while keeping all other parameters fixed, we can achieve excited state population
trapping for an optimum intensity of that laser. In Fig.3. we show this explicitly for an optimized intensity $ I_1 = 2$  W cm$^{-2}$. Note that at this laser intensity, PA stimulated linewidth $\Gamma_1$ is much larger than $\gamma_1$ meaning that the system is in the strong-coupling regime. In Fig. 4, we plot the dynamical behavior of the coherence $\rho_{12}$ as a function of $\gamma t$. It is clearly visible that the imaginary part is much more smaller than the real part. The upper panel of Fig. 4 shows that  Re$[\rho_{12}]$ becomes steady in the long time limit for an optimum frequency. Lower panel of Fig. 4 shows that Re$[\rho_{12}]$ becomes time-independent in the long time limit for the optimum parameters 
for which population in Fig. 3 becomes trapped. 

In conclusion, we have demonstrated that it is possible to generate and manipulate coherence between two excited ro-vibrational states of a molecule by using the technique of PA spectroscopy. A promising candidate for exploring such excited state coherence is  bosonic Yb atom. Once either or both the excited states are populated by PA, coherence between them builds up due to their interaction with the background electromagnetic vacuum. We have analyzed the effects of PA lasers on VIC. Our results show that under certain conditions population can be trapped in excited states. Note that VIC can be best realizable for an idealized three level system. We have discussed VIC in an atom-molecule interface system where one of the levels is the collisional contunuum of two ground state atoms. Our model relies on the condition that the two excited molecular levels decay to the same continuum. It may be very  hard to fulfill this condition in alkali metal atoms since they have several ground continua due to hyperfine interactions. Furthermore excited levels may decay to bound states in ground molecular configuration. Since bosonic Yb has no hyperfine interaction, it has only one ground continuum. Photoassociated excited rotational levels with large vibrational numbers as considered in this work are unlikely to decay to any bound level since their Frank Condon overlap with the least bound state close to the ground continuum is very small. Therefore, bosonic Yb appears to be a promising candidate for exploring VIC. The manipulation of VIC with PA may be important for controlling decoherence and dissipation in cold molecules. It may also be used for coherent control of atom-molecule conversion and optical  \cite{fedichev,jisha} and magneto-optical Feshbach resonance \cite{bdeb2}.  

AR gratefully acknowledge support from CSIR, Government of India. 



\begin{thebibliography}{999}
\bibitem{mandel} L. Mandel and E. Wolf {\it Optical Coherences and Quantum Optics}, Cambridge University Press, (1995).
\bibitem{ficek} Z. Ficek and S. Swain {\it Quantum Interference and Coherence}, Springer New-York, (2007).
\bibitem{fano} U. Fano, Phys. Rev. {\bf 124}, 1866 (1961).
\bibitem{miroshnichenko} A. E. Miroshnichenko, S. Flach and Y. S. Kivshar, Rev. Mod. Phys. {\bf 82} 2257 (2010).
\bibitem{agarwal} G. S. Agarwal, {\it Springer Tracts in Modern Physics: Quantum Optics} Springer-Verlag, Berlin, 19740.
\bibitem{gauthier} D. J. Gauthier, Y. Zhu and T. W. Mossberg, Phys. Rev. Lett. {\bf 66},  2460 (1991).
\bibitem{scully1} M. O. Scully and S. Y. Zhu, Science {\bf 281}, 1973 (1998). 
\bibitem{garraway}B. M. Garraway, M. S. Kim and P. L. Knight, Opt. Commun. {\bf 117}, 560 (1995).
\bibitem{zhou} P. Zhou and S. Swain, Phys. Rev. Lett. {\bf 77}, 3995(1996); 
Z. Ficek and S. Swain, Phys. Rev. A {\bf 69} 023401 (2004).
\bibitem{dutta}S. Dutta and K. Rai Dastidar, J. Phys. B:At. Mol. Opt. Phys. {\bf 40} 4287 (2007).
\bibitem{das08}S. Das, G.S. Agarwal and M.O. Scully, Phys. Rev. Lett. {\bf 101} 153601 (2008).
\bibitem{xia}H. R. Xia, C. Y. Ye and S. Y. Zhu, Phys. Rev. Lett. {\bf 77}, 1032 (1996); 
L. Li {\it et. al.} ibid. {\bf 84}, 4016 (2000).
\bibitem{Das10} S. Das and G. S. Agarwal, Phys. Rev. A {\bf 81}, 052341 (2010).
\bibitem{agarwal1} G.S. Agarwal, Phys. Rev. Lett.{\bf 84}, 5500(2000).
\bibitem{yang} Y. Yang, J. Xu, H. Chen and S. Zhu, Phys. Rev. Lett. {\bf 100}, 043601 (2008); 
J. Xu and Y. Yang, Phys. Rev. A, {\bf 81}, 013816 (2010).
\bibitem{evan}S. Evangelou, V. Yanopopas and E. paspalakis Phys. Rev. A {\bf 83}, 023819 (2011).
\bibitem{kiffner}M. Kiffner, J. Evers and C. H. Keitel, Phys. Rev. Lett. {\bf 96}, 100403 (2006).
\bibitem{das}S. Das and G.S. Agarwal, Phys. Rev. A {\bf 77}, 033850 (2008).
\bibitem{dutt} M.V.G. Dutt {\it et. al}, Phys. Rev. Lett. {\bf 94}, 227403 (2005). 
\bibitem{weiner} J. Weiner {\it et. al.},  Rev. Mod. Phys. {\bf 71}, 1 (1999). 
\bibitem{jones}K. M. Jones {\it et. al.},  Rev. Mod. Phys. {\bf 78}, 483 (2006).
\bibitem{bdeb1} B. Deb and G. S. Agarwal, J. Phys. B: At. Mol. Opt. Phys. {\bf 42}, 215203 (2009); 
B. Deb and A. Rakshit, {\it ibid} {\bf 42}, 195202 (2009).
\bibitem{bdeb2} B. Deb, J. Phys. B: At. Mol. Opt. Phys. {\bf 43}, 085208 (2010).
\bibitem{takahashi1} S. Tojo {\it et. al}, Phys. Rev. Lett. {\bf 96}, 153201 (2006).
\bibitem{takahashi2} M. Borkowski {\it et. al}, Phys. Rev. A {\bf 80}, 012715 (2009).
\bibitem{takahashi3} M. Kitagawa {\it et. al}, Phys. Rev. A {\bf 77}, 012719 (2008).
\bibitem{takahashi5} K. Enomoto {\it et.al.}, Phys. Rev. Lett. {\bf 98}, 203201 (2007)
\bibitem{takahashi4} K. Enomoto {\it et.al.}, Phys. Rev. Lett. {\bf 101}, 203201 (2008).
\bibitem{fedichev} P. O. Fedichev {\it et. al.}, Phys. Rev. Lett. {\bf 77}, 2913 (1996)
\bibitem{jisha} B. Deb and J. Hazra, Phys. Rev. Lett. {\bf 103}, 023201 (2009)
\end{thebibliography}
\end{document}